\documentclass[plb,preprint,tightenlines,floatfix,preprintnumbers,nofootinbib]{revtex4}
 \usepackage[dvips,final]{graphicx}
  \usepackage{amssymb}
   \usepackage{amsmath}
     \usepackage{epsfig}
     \usepackage{bm}
      \usepackage{pifont}
      \usepackage{simplewick}

\newcommand{\vecc}[1]{\mbox{\boldmath $#1$}}

\newcommand{\half}{\frac{1}{2}}

\newcommand{\be}{\begin{equation}}
\newcommand{\ee}{\end{equation}}

\newcommand{\ba}{\begin{eqnarray}}
\newcommand{\ea}{\end{eqnarray}}
\newcommand{\bg}{\begin{gather}}
\newcommand{\foma}{\end{gather}}

\newcommand{\nn}{\nonumber}

\newtheorem{theorem}{Theorem}[section]
\newtheorem{lemma}[theorem]{Lemma}
\newtheorem{proposition}[theorem]{Proposition}
\newtheorem{corollary}[theorem]{Corollary}
\newtheorem{definition}[theorem]{Definition}
\newtheorem{example}[theorem]{Example}

\newcommand{\qed}{\nobreak \ifvmode \relax \else
      \ifdim\lastskip<1.5em \hskip-\lastskip
      \hskip1.5em plus0em minus0.5em \fi \nobreak
      \vrule height0.75em width0.5em depth0.25em\fi}

\newcommand{\bd}{\begin{definition}}
\newcommand{\ed}{\end{definition}}
\newcommand{\bl}{\begin{lemma}}
\newcommand{\el}{\end{lemma}}
\newcommand{\bt}{\begin{theorem}}
\newcommand{\et}{\end{theorem}}
\newcommand{\bc}{\begin{corollary}}
\newcommand{\ec}{\end{corollary}}
\newcommand{\bex}{\begin{example}}
\newcommand{\eex}{\end{example}}
\newcommand{\bp}{\begin{proposition}}
\newcommand{\ep}{\end{proposition}}
\newcommand{\baa}{\begin{align}}
\newcommand{\eaa}{\end{align}}
\usepackage[usenames,dvipsnames]{xcolor}
\usepackage{pgfplots}
\usepackage{tikz}
\usetikzlibrary{decorations.pathmorphing}
\usetikzlibrary{decorations.markings}
\usetikzlibrary{intersections,shapes.arrows}
\usepackage{float}
\usepackage{enumerate}
\usetikzlibrary{calc}
\usetikzlibrary{shapes,arrows}
\usepackage{graphicx}
\usepackage{pdfpages}
\usepackage{wrapfig}
\usepackage{pgfplots}
\usepackage{tikz}
\usetikzlibrary{trees}
\usetikzlibrary{decorations.pathmorphing}
\usetikzlibrary{decorations.markings}
\usetikzlibrary{shapes.arrows}
\usepackage{refstyle}
\usetikzlibrary{patterns}
\tikzset{
		wilsonline/.style={draw, double distance=1.5pt, postaction={decorate}, decoration={markings,mark=at position .5 with 				 {\arrow{>}}}},
		wilsonline2/.style={draw=NavyBlue, double distance=2pt, postaction={decorate}}
		}
\tikzset{
		photon/.style={decorate, decoration={snake}, draw=red},
		electron/.style={draw=blue, postaction={decorate},decoration={markings,mark=at position .55 with {\arrow[draw=blue]{>}}}},
		gluon/.style={decorate, draw=magenta,     decoration={coil,amplitude=4pt, segment length=5pt}}
		}
\definecolor{kyelloworange}   {RGB}{255, 210,  110}
\tikzset{
point/.style={minimum size=0pt, inner sep=0pt},
dot/.style={minimum size=1.5pt, inner sep=1.3pt,circle,draw=NavyBlue!95,fill=NavyBlue!95},
photon/.style={decorate, decoration={snake}, draw=yellow,very thick},
gluon/.style={decorate, draw=Maroon, very thick,  decoration={coil,amplitude=2pt, segment length=3pt}},
quark/.style={draw=blue,very thick, postaction={decorate},
	decoration={markings,mark=at position .55 with {\arrow[draw=blue]{>}}}},
antiquark/.style={draw=blue,very thick, postaction={decorate},
	decoration={markings,mark=at position .55 with {\arrow[draw=blue]{<}}}},
eikonal/.style={double,double distance=1.5pt,draw=NavyBlue!60!blue, postaction={decorate},
	decoration={markings,mark=at position .55 with {\arrow[draw=NavyBlue!70!blue]{>}}}},
antieikonal/.style={double,double distance=1.5pt,draw=NavyBlue!60!blue, postaction={decorate},
	decoration={markings,mark=at position .55 with {\arrow[draw=NavyBlue!70!blue]{<}}}},
higgs/.style={draw=black,very thick, postaction={decorate},
	 decoration={markings,mark=at position .55 with {\arrow[draw=red]{>}}}}
}

\pgfkeys{
 	/marker/.is family, /marker,
	default/.style = {color=LimeGreen,inner sep=1.5mm},
	default2/.style = {color=LimeGreen,inner sep=0.5pt},
	color/.estore in = \markercolor,
	inner sep/.estore in = \markerinnersep
}
\newcommand{\Wilsonloop}{	
	\draw[double,double distance=1.5pt,draw=NavyBlue!70!blue] (0,0) -- (1,2) -- (3,2) -- (2,0) -- cycle;
}

\newcommand{\Looparrows}{	
	\path[very thick, postaction={decorate},
	decoration={markings,mark=between positions 0.15 and 1 step .25 with {\arrow[draw=NavyBlue!70!blue]{>}}}] (0,0) -- (1,2) -- (3,2) -- (2,0) -- cycle;
}
}] (0,0) -- (0,2) -- (2,2) -- (2,0) -- cycle;
}
}] (0,0) -- (2,0) -- (3,2) -- (1,2) -- cycle;
}

}] (0,0) -- (2,0) -- (2,2) -- (0,2) -- cycle;
}

\newcommand{\Looppoints}{
	\node[dot] at (0,0) {};
	\node[dot] at (1,2) {};
	\node[dot] at (3,2) {};
	\node[dot] at (2,0) {};
}

\newcommand\lr[1]{{\left({#1}\right)}}

\def\pd{\partial}
\def\pdlm{\partial_\mu}
\def\pdln{\partial_\nu}

\def\e{\epsilon}

\def\w{\omega}

\def\pd{\partial}

\def\<{\langle}
\def\>{\rangle}

\def\a{\alpha}

\def\g{\gamma}  \def\G{\Gamma}
\def\d{\delta}  
\def\l{\lambda}   
\def\s{\sigma}
\def\r{\rho}

\def\m{\mu}
\def\n{\nu}

\def\w{\omega}

\def\({\left(}
\def\[{\left[}
\def\){\right)}
\def\]{\right]}

\def\pd{\partial}

\usepackage{amsmath,tkz-euclide,tkz-fct}
\begin{document}

\title{Fr\'echet Derivative for Light-Like Wilson Loops}
\author{I.O.~Cherednikov}
\email{igor.cherednikov@uantwerpen.be}
\affiliation{EDF, Departement Fysica, Universiteit Antwerpen, B-2020 Antwerpen, Belgium}
\author{T.~Mertens}
\email{tom.mertens@uantwerpen.be}
\affiliation{EDF, Departement Fysica, Universiteit Antwerpen, B-2020 Antwerpen, Belgium}

\begin{abstract}
We address the equations of motion for the light-like QCD Wilson exponentials defined in the generalized loop space. We attribute an important class of the infinitesimal shape variations of the rectangular light-like Wilson loops to the Fr\'echet derivative associated to a diffeomorphism in loop space what enables the derivation of the law of the classically conformal-invariant shape variations. We show explicitly that the Fr\'echet derivative coincides (at least in the leading perturbative order) with the area differential operator introduced in the previous works. We discuss interesting implications of this result which will allow one to relate the rapidity evolution and ultra-violet evolution of phenomenologically important quantum correlation functions (such as 3-dimensional parton distribution functions) and geometrical properties of the light-like cusped Wilson loops.
\end{abstract}
\maketitle

\vspace{.4cm}

\section{Introduction}

 Quadrilateral planar Wilson loop with light-like sides \cite{WL_LC_rect_1,WL_LC_rect_2,WL_LC_rect_3} can be considered as a ``hydrogen atom'' of the Wilson loop theory in generalized loop space. Wilson loops having cusps and light-like segments show more complex renormalization and conformal properties than smooth and/or fully off-light-cone functionals. Analysis of the geometrical and dynamical properties of the generalized loop space, which can include cusped light-like Wilson exponentials, will deliver important information on the renormalization properties and evolution of various gauge-invariant quantum correlation functions, such as transverse-momentum dependent quark and gluon densities, multi-gluon scattering amplitudes, jet quenching parameter, etc.  (see, e.g., Refs. \cite{KR87,LargeX_KM,Makeenko_LC_WL,CS_1, CS_2, CS_3, CS_4, CKS_1,LC_TMD_1,LC_TMD_2,BMM_1,BM_1,CMTVDV_2013,Jet_1,Jet_15,Jet_2,Jet_3,Caputa:2012pi,Jet_4,Jet_5,Jet_6} and Refs. therein).

In the generalized loop space, the laws of ``motion'' are naturally formulated in terms of integro-differential equations for the Wilson loop which undergo certain variations of the underlying contours on which these path-ordered exponentials of the gauge fields are defined.  The infinitesimal local variations of the contours give rise to the variations of the exponentials themselves, the latter being described by the infinite set of the Makeenko-Migdal loop equations \cite{MM_WL_1,MM_WL_2,WL_Renorm_1,WL_Renorm_2,St_1,St_Kr_WL_cast_1,St_Kr_WL_cast_2}. On the other hand, physically meaningful transformations of the cusped light-like paths constitute a special class of motions in the generalized loop space which is not grasped straightforwardly by the Makeenko-Migdal approach. In this paper we show that the nonlocal area derivative of a Wilson loop which has been proposed in \cite{Cherednikov:2012yd,Cherednikov:2012ym,Cherednikov:2012qq,VanderVeken:2012kw,Mertens:2013xga} can be (at least in the lowest order of perturbative expansion) mathematically correctly introduced as a Fr\'echet derivative associated to a diffeomorphism with specific choice of the generating variational vector field in a generalized loop space setting (for details see Ref. \cite{DeG_2014,Tavares:1993pw} and Refs. therein).

The paper is organized as follows. In Section \ref{sec:frechet} we formally introduce the Fr\'echet derivative and recapitulate some of the results from Ref. \cite{Tavares:1993pw} to show how it links to diffeomorphisms with associated variational vector field. In Section \ref{sec:frechetwl} we apply this derivative to generic parallel transporters and  Wilson loops. In Section \ref{sec:frechet_lo}  we address the derivative on a specific Wilson loop, the light-like quadrilateral, and show that the leading-order contribution, when taking vacuum expectation values, is consistent with our derivative from Ref. \cite{Cherednikov:2012yd}.

\section{Fr\'echet derivative: Mathematical preliminaries}\label{sec:frechet}

In this Section we briefly outline necessary mathematical principles which allow us to  consistently define the objects under consideration.
The Fr\'echet derivative of an element $F$ in the generalized Wilson loop space is defined through the limit \cite{Lang:1979}
				\be
					\lim_{\Delta \to 0} \frac{\left\Vert F(x + \Delta) - F(x) - A \cdot \Delta \right\Vert_Y}{\left\Vert \Delta \right\Vert_X} = 0 \ ,
				\ee
where $\left\Vert ... \right\Vert_{X,Y}$ stands for the norm in a given space.
			If this limit exist, one says that
			\be
D F(x) = A
\ee
			is the {\it Fr\'echet derivative}\footnote{To be more accurate mathematically, we assume that $X,Y$ are Banach spaces and a function
			$F (x)$ exists in the subset $U \to Y$,
			and there
			exists a bounded linear operator
			$A: \ X \to Y$. For mathematical details and references to the original works, see \cite{DeG_2014}.}

Now we introduce the Chen iterated integrals \cite{chen1968,chen1954,chen1958,chen1971} which are defined as an iterative extension of the usual line integrals
		\be
			X(\g) = I_{i_1\cdots i_p}(\g) =
			\int_a^b I_{i_1\cdots i_{p-1}}(\g^t)\ dx_{i_p}(t) \ ,
		\ee
where $\g$ denotes a path (integration contour) in the generalized path/loop space.
After parametrization of the path $\gamma$ this becomes\footnote{In generalized loop space
	we assume reparametrization invariance, see also \cite{Makeenko_Methods} for a detailed discussion.}:
		\be
			X^{\w_1\cdots \w_r}(\g) =
			\int_\g \w_1 \cdots \w_r  = \int_0^1\left(\int_{\g^t} \w_1 \cdots \w_{r-1} \right)\w_r(t)dt \ ,
		\ee
	where $\w_k(t)\equiv \w_k(\g(t))\cdot\dot{\g}(t)$ and $\g^t$ represents the part of the path for $t\in [0,1]$.
	Note that the operators $\w_i$ are path-ordered under the integration, which will absorb the path-ordering operator $\mathcal{P}$ when considering Wilson loops in what follows.

	Considering now the generalized loop $\g\in \widetilde{LM}_p$ at the point $p$, with tangent space $T_{\g}{\cal LM}_p$ to $\widetilde{LM}_p$ at $\g$ which consists
	of sections of the pull-back bundle $\gamma^{*}T{\cal M}$. Put otherwise, it consists of the vector fields along $\gamma$, that vanish on $p$. Now choose a tangent vector
	\be
		V \in T_{\gamma}{\cal PM}_p \ ,
	\ee	
	  and let $s \mapsto \gamma_s$ be
	a curve of paths in ${\cal PM}_p$, starting at $\gamma$.
	We have then
	\be
	V(t) =
\left. \frac{\pd}{\pd s}\right\vert_{s=0} \g_s(t) \  ,
	\ee
	from now on referred to as the {\it variational vector field}.

	In Ref. \cite{Tavares:1993pw}, Tavares shows that the Fr\'echet derivative of $X^{\w_1\cdots \w_r}(\g)$ at $\g$ can be written as follows
		\ba
			A_\g & = & D_{V}\  X^{\w_1...\w_r}(\g) = \sum _{i=1}^{r}
									\int_{\g}\w_1...\w_{i-1} \cdot {{\cal J}_{V}}(d\w_i)\cdot \w_{i+1}...\w_r \nn\\
								 &+& \sum_{i=2}^{r} \int_{\g}\w_1...\w_{i-2} \cdot {\cal J}_{V}(\w_{i-1} \wedge \w_i) \cdot \w_{i+1}...\w_r
								+  \left(\int_{\g}\w_1...\w_{r-1}\right) \cdot \w_r(V(1)) \ ,
		\ea
	where for a closed path $V(0)=V(1)=0$ and ${\cal J}_V$ is defined as the interior product \cite{frankel2011geometry}
\be
	{\cal J}_V: \bigwedge^{p}(M)\to \bigwedge^{p-1}(M),
\ee
	with $M$ a differentiable manifold and defined by:
		\begin{subequations}
			\begin{alignat}{3}
				&{\cal J}_V \a^0 &&= 0, \ \ \ \ \ &&\text{if $ \a^0$ is a 0-form},\\
				&{\cal J}_V \a^1 &&= \a(V), \ \ \ \ \ &&\text{if $ \a^1$ is a 1-form},\\
				&{\cal J}_V \a^p  (w_2,\cdots ,w_p) &&= \a(V,w_2,\cdots ,w_p) , \ \ \ \ \ &&\text{if $ \a^p$ is a p-form}.
			\end{alignat}
		\end{subequations}
Therefore one obtains
		\ba\label{eq:frechetdertavares1}
			D_{V} \ X^{\w_1...\w_r}(\g) &=& \sum _{i=1}^{r}
									\int_{\g}\w_1...\w_{i-1} \cdot {{\cal J}_{V}}(d\w_i)\cdot\w_{i+1}...\w_r \nn\\
								 &+& \sum_{i=2}^{r} \int_{\g}\w_1...\w_{i-2} \cdot {\cal J}_{V}(\w_{i-1} \wedge \w_i) \cdot \w_{i+1}...\w_r \ .
		\ea		
If one restricts the variational vector field $V$ to be induced by a vector field ${Y} \in {\cal X}_p{\cal M}$, i.e., $V={Y} \circ \gamma$ (for example, if $\gamma$ is embedded), then we observe that the Fr\'echet derivative coincides with the derivative associated with a diffeomorphism of the manifold $M$ that is infinitesimally generated by the vector field $Y$, see Ref. \cite{Tavares:1993pw}.

\section{Fr\'echet derivative of a Wilson loop}
\label{sec:frechetwl}	

We define a {\it Wilson loop} ${\cal W_\gamma}$ as a vacuum average of the traced operator-valued exponential
\be
	{\cal U}_{\g^t} = \exp{\left[\int\limits_{\g}^t\! {\cal A}_\m(x)\ dx^\m \right]} \ ,
	\label{eq:paralleltransporter_0}
\ee
where ${\cal A}$ belongs to the Lie algebra of the gauge group $SU(N_c)$, that is
\be
{\cal W}_{\gamma}
=
\Big\langle 0 \Big| \frac{1}{N_c} {\rm Tr} \ {\cal U}_\g \Big| 0 \Big\rangle \ .
\ee
Applying the operation (\ref{eq:frechetdertavares1}) to the parallel transporter (\ref{eq:paralleltransporter_0}), one obtains for the logarithmic Fr\'echet derivative \cite{Tavares:1993pw}
		\be\label{eq:frechetder1}
			D_V [{\cal U}_\g]
			=
			{\cal U}_\g \cdot \int\limits_0^1\! dt \ {\cal U}_{\g^t}\cdot {\cal F}_{\m\n}(t) \[V^\m(t)\wedge \dot{\g}^\n(t) \] \cdot {\cal U}_{\g^t}^{-1} \ .
		\ee	
where ${\cal U}_{\g^t}$ is interpreted now as the operator-valued parallel transporter (see also Eq. (\ref{eq:paralleltransporter_0})) along the part of the path $\gamma$ from the point $0$ to $t$, and the vector field $V$, associated with the diffeomorphism flow, determines the direction of the variation of the loop.	

	From Eq. (\ref{eq:frechetder1}) it is now clear that this derivative is closely related to the area derivative of the parallel transporter
	around a loop $\g$:
		\be
			\triangle^E_{(\epsilon; \ u \wedge v)}(p)\ {\cal U}_{\g} =   {\cal U}_{\g}\cdot {\cal F}_{\m\n} (u^\m \wedge v^\n) \  ,
		\ee
	which depends on the two independent vector fields $\{u,v\}$ and where $ {\cal F}_{\m\n}$
	is the usual field strength tensor (or curvature tensor), by taking one of the vector fields
	to be the tangent to the loop and integrating over it along the loop.
	\begin{figure}[h]
	\centering
	\begin{minipage}[b]{.4\textwidth}
			\centering
			\hspace*{-2.5cm}%
			\begin{tikzpicture}[allow upside down,scale= .7]
			\draw[black,line width=1pt] (0,0).. controls +(right:6cm) and +(left:4cm) .. (2,6)
				\foreach \p in {0,5,...,100} {
					node[sloped,inner sep=0cm,above,pos=\p*0.01,
					anchor=south west, minimum height=(10+\p)*0.03cm,
					minimum width=(10+\p)*0.03cm] (N \p){}
					}
				     	.. controls +(right:2cm) and +(right:3cm) .. (6,2)
			    	\foreach \p in {5,10,...,100} {
			      		node[sloped,inner sep=0cm,above,pos=\p*0.01,
				      	anchor=south west,
			      		minimum height=(110-\p)*0.03cm,minimum width=(110-\p)*0.03cm]
			      		(N2 \p){}
			    		}
			    		..controls +(left:6cm) and +(right:4cm) .. (1,-2)
			      	\foreach \p in {5,10,...,100} {
			      		node[sloped,inner sep=0cm,above,pos=\p*0.01,
			      		anchor=south west,
			      		minimum height=(110-\p)*0.03cm,minimum width=(110-\p)*0.03cm]
			      		(N3 \p){}
			    		}
			    		.. controls +(left:2cm) and +(left:3cm) .. (0,0)
			      	\foreach \p in {5,10,...,100} {
			      		node[sloped,inner sep=0cm,above,pos=\p*0.01,
			      		anchor=south west,
			      		minimum height=(110-\p)*0.03cm,minimum width=(110-\p)*0.03cm]
			      		(N4 \p){}
			    		}
			    		;
			\filldraw[draw=Maroon,line width=2pt,fill=gray,pattern=north east lines] (2,6.5) -- (2.5,6.5) -- (2.5,6) -- (2,6)-- cycle;
			\draw[-latex,blue,line width=1pt] (2,6) -- (2,9);
			\draw[-latex,color=green!50!black,line width=1pt] (2,6) -- (5,6);
		\end{tikzpicture}
		\vspace*{-3cm}
		\caption{Local area derivative.}
		\label{fig:areadervar}
		\end{minipage}
		\begin{minipage}[b]{.5\textwidth}
			\begin{tikzpicture}[allow upside down,scale=.7]
							\draw[black,line width=1pt] (0,0).. controls +(right:6cm) and +(left:4cm) .. (2,6)
							    	\foreach \p in {0,5,...,100} {
							     		node[sloped,inner sep=0cm,above,pos=\p*0.01,
									anchor=south west, minimum height=(10+\p)*0.03cm,
									minimum width=(10+\p)*0.03cm] (N \p){}
									}
							     		.. controls +(right:2cm) and +(right:3cm) .. (6,2)
							     	\foreach \p in {5,10,...,100} {
							      		node[sloped,inner sep=0cm,above,pos=\p*0.01,
							      		anchor=south west,
							      		minimum height=(110-\p)*0.03cm,minimum width=(110-\p)*0.03cm]
							      		(N2 \p){}
							    		}
							    		..controls +(left:6cm) and +(right:4cm) .. (1,-2)
							      	\foreach \p in {5,10,...,100} {
							      		node[sloped,inner sep=0cm,above,pos=\p*0.01,
							      		anchor=south west,
							      		minimum height=(110-\p)*0.03cm,minimum width=(110-\p)*0.03cm]
							      		(N3 \p){}
							    		}
							    		.. controls +(left:2cm) and +(left:3cm) .. (0,0)
							      	\foreach \p in {5,10,...,100} {
							      		node[sloped,inner sep=0cm,above,pos=\p*0.01,
							      		anchor=south west,
							     		minimum height=(110-\p)*0.03cm,minimum width=(110-\p)*0.03cm]
							     		(N4 \p){}
							    		}
							   		 ;
							    	\foreach \p in {0,5,...,100} {
							      		\draw[-latex,blue,line width=1pt] (N \p.south west) -- (N \p.north west);
							      		\draw[-latex,color=green!50!black,line width=1pt] (N \p.south west) -- (N \p.south east);
							    		}
							    	\foreach \p in {5,10,...,100} {
							      		\draw[-latex,blue,line width=1pt] (N2 \p.south west) -- (N2 \p.north west);
							      		\draw[-latex,color=green!50!black,line width=1pt] (N2 \p.south west) -- (N2 \p.south east);
							   		 }
			 				\draw[Maroon,line width=2pt] (0,0).. controls +(right:5.8cm) and +(left:5.8cm) .. (1.9,6.8).. controls +(right:2.8cm) and +(right:3cm) .. (6,2);
				  	\end{tikzpicture}
					\vspace*{-3cm}
		\caption{Fr\'echet derivative.}
		\label{fig:frechetdervar}
		\end{minipage}
	\end{figure}

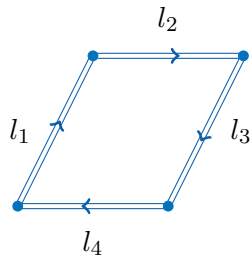
\begin{figure}[h]
		\begin{tikzpicture}
		\node[point] at (-1.3,0){};
		\begin{scope}[shift={(0,-0.75)}, scale=1]
			\Wilsonloop
			\Looppoints
			\Looparrows
			\node[left=.5em] at (.5,1) {$l_1$};
			\node[above=.5em] at (2,2) {$l_2$};
			\node[right=.5em] at (2.5,1) {$l_3$};
			\node[below=.5em] at (1,0) {$l_4$};
		\end{scope}
	\end{tikzpicture}
	\caption{Wilson loop parametrization.}
	\label{fig:parametrization}
	\end{figure}

	Figures \ref{fig:areadervar} and \ref{fig:frechetdervar} visualise the relation between the two
	derivatives, where the arrows represent the vector fields, where in Fig. \ref{fig:frechetdervar}
	one of the fields is tangent to the curve. Notice that in Fig. \ref{fig:frechetdervar}
	the small ``square'' formed between the original, the deformed curve and the ``normal''
	vector field arrows are actually pointed area derivatives (i.e. the area derivatives operating
	on specific points). Integration over these area derivatives then results in the deformed curve
	(the thick curve in Fig.\ref{fig:frechetdervar}). In the next Section we show that the derivatives
	
		\ba
			S_{12}\frac{\delta}{\delta S_{12}}  & = &
			 (2{\ell}_1 \cdot \ell_2) \frac{\delta}{\delta  (2\ell_1 \cdot \ell_2)} = \ell_1^{+}  \frac{\delta}{\delta \ell_1^{+}}\\
			S_{23}\frac{\delta}{\delta S_{23}}  & = &
			(2\ell_2 \cdot \ell_3) \frac{\delta}{\delta  (2\ell_2 \cdot \ell_3)} = \ell_2^{-}  \frac{\delta}{\delta \ell_2^{-}} \ ,
		\ea
	with $S_{ij}$ being the adapted Mandelstam-like variables associated with the Wilson loop (with the parametrization shown in Fig. \ref{fig:parametrization})
	defined in
	\cite{Cherednikov:2012yd,Cherednikov:2012ym,Cherednikov:2012qq,VanderVeken:2012kw} and used in \cite{Mertens:2013xga}
	are the lowest order contributions of the logarithmic Fre\'chet derivatives with the appropriate vector field $V^\m$ as
	generator for diffeomorphism transformation associated to the Fr\'echet derivative, as stated before.

\section{Calculation of the leading-order contributions}
\label{sec:frechet_lo}

	The perturbative expansion of the parallel transporter (\ref{eq:paralleltransporter_0}) written in terms of Chen iterated integrals
	\cite{chen1968,chen1954,chen1958,chen1971,Tavares:1993pw}
	is given by
		\be\label{eq:paralleltransporter}
			{\cal U}_\g := 1 + \int {\cal A}_\m(x) \ dx^\m + \int {\cal A}_\m(x) {\cal A}_\n(y) \ dx^\m dy^\n+\cdots \ ,
		\ee
	where  the operators ${\cal A}_\m \cdots {\cal A}_\n$ are ordered as defined by the Chen integrals. For the inverse path (with reverse ordering and correct sign) one has
		\be
			{\cal U}_{\g^{-1}} := 1 - \int {\cal A}_\m (x) \ dx^\m + \int {\cal A}_\n (y) {\cal A}_\m (x) \ dx^\m dy^\n-\cdots \ .
		\ee
Given that the non-Abelian field strength tensor reads
		\be
			{\cal F}_{\m\n}  = (d{\cal A})_{\m\n}  +  {\cal A}_\m \wedge {\cal A}_\n \ ,
		\ee
	we expand Eq. (\ref{eq:frechetder1}) to lowest non-trivial order:
		\ba
		 & &	D_V[{\cal W}_\g]_{{\rm LO}} = \nn \\
		 & & {\textbf{1} }\cdot\oint\limits_0^1dt\  \left[\left(
				\oint\limits_0^t {\cal A}_\s (x(s)) \ \frac{dx^\s}{ds} ds \cdot
				\left\lbrace\pdlm {\cal A}_\n(y(t))- \pdln {\cal A}_\m (y(t))\right\rbrace \left\lbrace V^\m(y(t))\wedge \dot{\g}^\n(y(t))\right\rbrace \cdot
				{\bf 1}
				\right)\right.\nn\\
				&&\ \ \ \ \
				\left.
				- \left(
				{\bf 1}\cdot
				\left\lbrace\pdlm {\cal A}_\n (y(t))- \pdln {\cal A}_\m (y(t))\right\rbrace \left\lbrace V^\m(y(t))\wedge \dot{\g}^\n(y(t))\right\rbrace \cdot
				\oint\limits_0^t {\cal A}_\l (x(u)) \ \frac{dx^\l}{du} du
				\right)
				\right]\nn\\
				&&\ \ \ \ \
				+\oint\limits_0^1 {\cal A}_\s (x) \frac{dx^\s}{ds} ds \cdot  \oint\limits_0^1 dt \ {\bf 1} \cdot
				\left\lbrace\pdlm {\cal A}_\n (y(t))- \pdln {\cal A}_\m (y(t))\right\rbrace \left\lbrace V^\m(y(t))\wedge \dot{\g}^\n(y(t))\right\rbrace \cdot
				{\bf 1} \ ,\nn\\
		\label{eq:LO_1}
		\ea
	where the term with the minus in the first contribution originates from the inverse path.	
	Calculating the vacuum expectation value of the r.h.s. of Eq. (\ref{eq:LO_1}), we have to Wick contract the different fields in the factors and terms to
	acquire the propagators. It is worth remarking that the partial derivatives $\pdlm,\pdln$
	are defined with respect to the coordinate $y$, i.e.
	$$
	\pdlm=\frac{\pd}{\pd y^\m}\ , \ \pdln=\frac{\pd}{\pd y^\n} \ .
	$$
	Due to the path reduction property the lowest order contribution in the first term cancels\footnote{Since the contributions have an opposite sign due to the inverse ordering on the inverse path.},
	what was also checked by explicit calculations using the coordinate expression for the gluon propagator in the
	Feynman gauge:
		\be
			\langle 0 | T [A_\mu^a (x)A_\nu^b (y) ] | 0 \rangle
=			
			 D_{\m\n}^{ab}(x-y)
			=
			\frac{(\m^2 \pi)^\e}{4\pi^2}\G(1-\e)\ \frac{g_{\m\n}\d^{ab}}{\left[-(x-y)^2\right]^{1-\e}} \ .
		\ee
	The cancelation of these terms is graphically represented in Figs. \ref{fig:frechetpathred} and \ref{fig:wlpathred}.
	As a result we only need to consider the Wick contractions of the second term of Eq. (\ref{eq:LO_1}) which are the following:
			 \be \contraction{}{ A}{_\s^b(x(\s))\pdlm}{A} \nomathglue{A_\s^a(x(\s))\pdlm A}{_\n^b(y(\s'))} = \pdlm D_{\s\n}^{ab}(x-y) = \delta^{ab}\pdlm D_{\s\n}(x-y) \label{wick1},\ee
			\be \contraction{}{ A}{_\s^b(x(\s))\pdln}{A} \nomathglue{A_\s^a(x(\s))\pdln A}{_\m^b(y(\s'))}= \pdln D_{\s\m}^{ab}(x-y) =\delta^{ab}\pdln D_{\s\m}(x-y)\label{wick2}.\ee
	\begin{figure}[h]
	\centering
	\begin{minipage}[b]{.4\textwidth}
		\begin{tikzpicture}[allow upside down,scale=.6,baseline]
			\centering
			\hspace{-3cm}
							\tkzDefPoint(0,0){0}
							\tkzDefPoint(2,6){a}
							\tkzDrawPoints(0,a)
							\tkzDefPoint(2,6.2){t}
							\tkzLabelPoints[above left](0,t)
							\draw[electron,black,line width=1pt] (0,0).. controls +(right:6cm) and +(left:4cm) .. (2,6)
							    	\foreach \p in {0,5,...,100} {
							     		node[sloped,inner sep=0cm,above,pos=\p*0.01,
									anchor=south west, minimum height=(10+\p)*0.03cm,
									minimum width=(10+\p)*0.03cm] (N \p){}
									}
							     		.. controls +(right:2cm) and +(right:3cm) .. (6,2)
							     	\foreach \p in {5,10,...,100} {
							      		node[sloped,inner sep=0cm,above,pos=\p*0.01,
							      		anchor=south west,
							      		minimum height=(110-\p)*0.03cm,minimum width=(110-\p)*0.03cm]
							      		(N2 \p){}
							    		}
							    		..controls +(left:6cm) and +(right:4cm) .. (1,-2)
							      	\foreach \p in {5,10,...,100} {
							      		node[sloped,inner sep=0cm,above,pos=\p*0.01,
							      		anchor=south west,
							      		minimum height=(110-\p)*0.03cm,minimum width=(110-\p)*0.03cm]
							      		(N3 \p){}
							    		}
							    		.. controls +(left:2cm) and +(left:3cm) .. (0,0)
							      	\foreach \p in {5,10,...,100} {
							      		node[sloped,inner sep=0cm,above,pos=\p*0.01,
							      		anchor=south west,
							     		minimum height=(110-\p)*0.03cm,minimum width=(110-\p)*0.03cm]
							     		(N4 \p){}
							    		}
							   		 ;
							    	\foreach \p in {100} {
							      		\draw[-latex,blue] (N \p.south west) -- (N \p.north west);
							      		\draw[-latex,color=green!50!black] (N \p.south west) -- (N \p.south east);
							    		}
								\node[below=.1cm] at (N 100.south east) {${\color{green!50!black} \dot{\g}(t)}$};	
								\node[right=.1cm] at (N 100.north west) {${\color{blue} V^\m(t)}$};	
							\draw[electron,red,line width=.5 pt] (0,0.2).. controls +(right:5.5cm) and +(left:4.5cm) .. (2,6.2);
							\draw[electron,blue,line width=2 pt] (2,5.8).. controls +(left:3.5cm) and +(right:6.5cm) .. (0,-0.2);
							\node[above=.15cm] at (0,0) {${\color{red} {\g}^t}$};	
							\node[below=.15cm] at (0,0) {${\color{blue} {\left(\g^t\right)}^{-1}}$};
				  		\end{tikzpicture}
						\vspace{-3cm}
						\caption{Fr\'echet path reduction.}
						\label{fig:frechetpathred}
		\end{minipage}
		\begin{minipage}[b]{.4\textwidth}
			\begin{tikzpicture}[scale=1.5]
				\node[point] at (-1.3,0){};
				\begin{scope}[shift={(0,-0.75)}, scale=1]
					\Wilsonloop
					\Looppoints
					\Looparrows
					\node[left=.5em] at (.5,1) {$l_1$};
					\node[below=.3em] at (2,2) {$l_2$};
					\node[right=.5em] at (2.5,1) {$l_3$};
					\node[above=.3em] at (1,0) {$l_4$};
					\draw (0,0) node[anchor=north] {$x_1=\g(0)$};
					\draw (1,2) node[left] {$x_2$};
					\draw (3,2) node[anchor=south] {$x_3$};
					\draw (2,0) node[anchor=north] {$x_4$};
					\draw[electron,red,line width=.5 pt] (-.2,0) -- (.9,2.2) -- (1.5,2.2);
					\draw[electron,blue,line width=2 pt] (1.5,1.8) -- (1.1,1.8) -- (.2,0);
					\node[dot] at (1.5,2) {};
					\node[label={[label distance=.1em]60:$\g(t)$}] at (1.5,2) {};
					\draw[-latex,blue,line width=1pt] (1.5,2) -- (1.5,4);
					\draw[-latex,green!50!black,,line width=1pt] (1.5,2) -- (4,2);
					\node[right=.1cm] at (1.5,4) {${\color{blue} V^\m(t)}$};
					\node[below=.1cm] at (4,2){${\color{green!50!black} \dot{\g}(t)}$};
				\end{scope}
		\end{tikzpicture}
		\caption{Path reduction for WL.}
		\label{fig:wlpathred}
		\end{minipage}
	\end{figure}




	Before starting the explicit calculation of these remaining contributions we have to choose an appropriate vector field $V^\m$
	that will generate the same deformation as the $S_{12}\frac{\delta}{\delta S_{12}}$ from
	\cite{Cherednikov:2012yd}.
	Choosing $V^\m := ({\ell}_1^{+} \sigma, 0^{-}, \vecc 0_{\perp}) \ , \  \sigma \in [0,1]$ we see immediately
	that this will restrict the possible contributions from the wedge product $V^\m \lr{y\lr{\sigma}}\wedge \dot{\g}^\n\lr{y\lr{\sigma}}$
	due to its anti-symmetric nature:

		\begin{itemize}
\label{list1}

			\item {Along ${\ell}_1$: $V^\mu \wedge \dot{\g}^\nu =0 $, what follows from the asymmetry of the wedge product and the fact that both vectors are parallel}

			\item {Along ${\ell}_2$: $V^\mu \wedge \dot{\g}^\nu = -\ell_1^{+} \ell_2^{-}(\pd_{+}\wedge \pd_{-})$, due to (anti-)linearity of the wedge product}

			\item {Along ${\ell}_3$: $V^\mu \wedge \dot{\g}^\nu =0 $, what follows from the asymmetry of the wedge product and the fact that both vectors are parallel}

			\item {Along ${\ell}_4$: $V^\m \wedge \dot{\g}^\n =0 $, because we assume the vector field to be zero along the part of the path.}

		\end{itemize}
	Combining the above restrictions with the remaining Wick contractions shown in Eq. (\ref{wick1}) and Eq. (\ref{wick2}) it is easy to see that each of the contractions gives rise to four terms, one for each side of the quadrilateral in Fig. \ref{fig:parametrization} so that we end up with a total of eight
	terms which we calculate below. Notice that due to the above restrictions, in the remaining Wick contractions, $y$ is restricted to the top line in the diagram shown in Fig. \ref{fig:parametrization}.
	
		\subsection{$\pdlm D_{\s\n}(x-y)-\pdln D_{\s\m}(x-y)$ term with $x\in \ell_1$}
		
			Parametrizing the paths for $x$ and $y$ as (assuming that $x_1 = 0$):
				\ba
					x &=& \sigma \ell_1 ,\ \sigma \in [0,1] \\
					y &=& \ell_1 + \sigma' \ell_2,\ \sigma'\in [0,1] \ ,
				\ea
			we have:
				\ba
					dx^\s &=& \left(\frac{dx^\s}{d\sigma}\right)d\sigma=(\ell_1^{+}, 0^{-}, \vecc 0_\perp)d\sigma\nn\\
			dy^\n &=& \left(\frac{dy^\n}{d\sigma'}\right)d\sigma'=(0^{+},\ell_2^{-}, \vecc 0_\perp)d\sigma'=\dot{\g}(\sigma')d\sigma'\nn\\
					x-y &=& (\sigma - 1) \ell_1 - \sigma' \ell_2\nn\\
					(x-y)^2 &=& -2(\sigma-1)\sigma' \ (\ell_1^{+}\ell_2^{-}) \ .\nn
				\ea
			For notational simplicity let us define:
				\be
					K_\e :=  \frac{(\m^2 \pi)^\e}{4\pi^2}\G(1-\e)\  .
				\ee	
			Calculating this contribution:
				\ba\label{goal1}
					&&\int\limits_0^1\ d\s'\ d\s\frac{dx^\r}{d\s}
						\left(\frac{\pd}{\pd y^\m} D_{\r\n}(x-y) - \frac{\pd}{\pd y^\n} D_{\r\m}(x-y)\right)
						\left[V^\m(y) \wedge \dot{\g}^\n(y)\right]\nn\\
										&=& 	
							K_\e \int\limits_0^1\ d\s'\ d\s\frac{dx^\r}{d\s}
								\left[
									\left(\frac{dy^\n}{d\s'}\frac{2(\e-1)g_{\r\n} (x-y)_\m V^\m(\s')}{\left[-(x-y)^2\right]^{2-\e}}\right)
									-
									\left(\frac{dy^\n}{d\s'}\frac{2(\e-1)g_{\r\m} (x-y)_\n V^\m(\s')}{\left[-(x-y)^2\right]^{2-\e}}\right)
								\right]
							\nn\\	
						&=&
							K_\e
							\left[
								\left(\frac{(1-\e)}{2}(- S_{12})^\e \int\limits_0^1 \  \frac{d\s\ d\s'}{\s'^{1-\e}(\s-1)^{2-\e}}\right)
								-
								\left(\frac{(1-\e)}{2}(- S_{12})^{\e-1}(\ell_1)^2 \int\limits_0^1 \  \frac{d\s\ d\s'}{\s'^{1-\e}(\s)^{2-\e}}\right)
							\right]	
							\nn\\
						&=&
							\half K_\e \frac{S_{12}^\e}{\e}\label{result1_4_1} \ ,
				\ea
			where $S_{ij}$ represents the Mandelstam-like variable for the pair of vectors $\ell_{i,j}$.
			Which is exactly the same result as taking the derivative $\ell_1\frac{ \delta}{\delta \ell_1}$ of the
			original integral:
				\be
					\ell_1 \frac{\delta}{\delta \ell_1}K_\e \oint\! \frac{g_{\m\n}\ dx^\m\ dy^\n}{\left(-(x-y)^2\right)^{1-\e}}
					=
					\ell_1 \frac{\delta}{\delta \ell_1}K_\e \oint\! \frac{(\ell_1 \ell_2)\ d\s\ d\s'}{\left(-(2\ell_1 \ell_2(\s - 1) \s')^2\right)^{1-\e}} =
					\half K_\e \frac{S_{12}^\e}{\e} \  .
				\ee
		
		\subsection{$\pdlm D_{\r\n}(x-y)-\pdln D_{\r\m}(x-y)$ term with $x\in \ell_2$}
			
			This term is trivial since it reduces to a self-energy on the light-cone which in dimensional
			regularization is formally zero.
		
		\subsection{$\pdlm D_{\r\n}(x-y)-\pdln D_{\r\m}(x-y)$ term with $x\in \ell_3$}	
			
			Making use of the symmetry $2\ell_1 \ell_2 = - 2 \ell_2 \ell_3 = S_{23}$, where now $S_{23}$ is
			the second Mandelstam variable we can write down this contribution immediately:
				\be
					\int\limits_0^1\ d\s'\ d\s\frac{dx^\r}{d\s}
						\left(\frac{\pd}{\pd y^\m} D_{\r\n}(x-y) - \frac{\pd}{\pd y^\n} D_{\r\m}(x-y)\right)
						\left[V^\m(y) \wedge \dot{\g}^\n(y)\right] = 0 \ .
				\ee
			which is again the same as taking the derivative  $\ell_1\frac{ \delta}{\delta \ell_1}$ since the
			original integral is formally independent of $\ell_1$ thus resulting in zero.
		
		\subsection{$\pdlm D_{\r\n}(x-y)-\pdln D_{\r\m}(x-y)$ term with $x\in \ell_4$}	
			
			This contribution is actually the most tricky to calculate, where the intricacies
			of the calculation are hidden in the combination of the integration and derivatives
			with respect to $y$. So here we will apply a slightly different approach
			then in the derivations above. Instead of evaluating the integrals
			we will keep the integrals and show that the taking the derivative $\ell_1\frac{ \delta}{\delta \ell_1}$
			results in the same integrals as when we take the Fr\'echet derivative.
			Using the parametrization:
				\ba
					x &=& -(1-\s) \ell_4 ,\ \sigma \in [0,1] \ , \\
					y &=& \ell_1 + \s' \ell_2 \ ,
					\ \sigma'
				\in
				[0,1] \ Ê,
				\ea
			we start by splitting up the calculations in the
			contributions $\pdlm D_{\r\n}(x-y)$ and $-\pdln D_{\r\m}(x-y)$.
			For the first term $\pdlm D_{\r\n}(x-y)$ we proceed as before resulting in:
				\ba\label{eq:result2}
					& & \int\limits_0^1\ d\s'\ d\s\frac{dx^\r}{d\s}
						\left(\frac{\pd}{\pd y^\m} D_{\r\n}(x-y)\right)
						\left[V^\m(y) \wedge \dot{\g}^\n(y)\right]
						=\nn\\
							&&-2 K_\epsilon (\epsilon -1) \int\limits_0^1\ d\s'\ d\s\ \left[\ell_1\cdot (\ell_1 + \s' \ell_2 + (1-\s) \ell_4) \right]\frac{(\ell_2\cdot \ell_4)}
							{\left(-(\ell_1+\s'\ell_2+(1 - \s)\ell_4)^2\right)^{2-\e}} \ ,\nn\\
				\ea
			the second term is the tricky one.
			If we look at the index of the derivative with respect to $y$ (i.e. $\n$)
			one can see that then afterwards we integrate again over $dy^\n$, so
			that we might as well evaluate the original kernel $\frac{1}{\left(-(x-y)^2\right)^{1-\e}}$ between
			its boundary values as one would do by a normal integration. This results in:
				\ba
					& & - \int\limits_0^1\ d\s'\ d\s\frac{dx^\r}{d\s}
						\left( \frac{\pd}{\pd y^\n} D_{\r\m}(x-y)\right)
						\left[V^\m(y) \wedge \dot{\g}^\n(y)\right]
						=\nn\\
						&&
							- K_\epsilon\int\limits_0^1\ d\s
							(\ell_1\cdot \ell_4) \s'
							\left[
								\frac{1}{\left(\ell_1 + \ell_2 + (1 - \s') \ell_4\right)^{2(1-\e)}}
								-
								\frac{1}{\left(\ell_1 + \s' \ell_4\right)^{2(1-\e)}}
							\right] = 0 \ ,\nn\\
				\ea	
			where we used $(\ell_2 \cdot \ell_4 ) = 0$  and $\ell_1\ell_2 = - \ell_1 \ell_4$ making the two integrals equal which of course after
			subtraction results in the zero.
			Taking the $\ell_1\frac{ \delta}{\delta \ell_1}$ of the original integral results in:
				\ba
					& & \ell_1\frac{ \delta}{\delta \ell_1} \int\limits_0^1\ d\s'\ d\s\frac{dx^\r}{d\s}\frac{dy^\m}{d\s'}
						\left( D_{\r\m}(x-y)\right)=\nn\\
					& & - 2 K_\epsilon (\epsilon -1) \int\limits_0^1\ d\s' \ d\s\
					\left[\ell_1\cdot (\ell_1 + \s' \ell_2 + (1 - \s) \ell_4) \right]\frac{(\ell_2\cdot \ell_4)}{\left(-(\ell_1 + \s' \ell_2 + (1 - \s) \ell_4)^2\right)^{2-\e}} \ ,
				\ea
			which is the same as Eq. (\ref{eq:result2}) as desired.
			
			Similar calculations with the variational vector field now chosen $(0^{+}, \ell_2^{-},\vecc 0_\perp)$ and
			the point $y$ restricted to the side $\ell_3$ of the quadrilateral (due to the anti-symmetry of the wedge product)
			result in the contribution:
				\be
					\half K_\e \frac{S_{23}^\e}{\e}-2 (\epsilon -1) \int\limits_0^1\ d\s'\ d\s\ \left[\ell_4\cdot (\ell_4 + \s' \ell_1 + (1-\s) \ell_3)
					\right]\frac{(\ell_1\cdot \ell_3)}{\left(-(\ell_4 + \s' \ell_1 + (1 - \s) \ell_3)^2\right)^{2-\e}} \ ,
				\ee
			with $S_{23}=2(\ell_2 \cdot \ell_3)$.
			
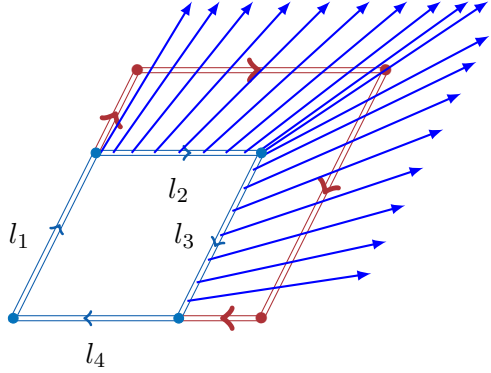
\begin{figure}[h]
			\begin{tikzpicture}[allow upside down,scale=1.1]
					\begin{scope}[shift={(0,-0.75)}, scale=1]
						\Wilsonloop
						
						\Looparrows
						\node[left=.5em] at (.5,1) {$l_1$};
						\node[below=.5em] at (2,2) {$l_2$};
						\node[left=.5em] at (2.5,1) {$l_3$};
						\node[below=.5em] at (1,0) {$l_4$};
						
						\path (0,0)--node[left=.5em] {} (1,2);
						\path(1,2).. controls +(right:0cm) and +(left:0cm) .. (3,2)
						    	\foreach \p in {0,10,...,100} {
						     		node[sloped,inner sep=0cm,above,pos=\number\p*0.01,
								anchor=south west, minimum height=(10)*0.2cm,
								minimum width=(\p)*0.02cm+1cm] (N \p){}
								};
						\path (3,2).. controls +(right:0cm) and +(left:0cm) .. (2,0)
							\foreach \p in {0,10,...,100} {
						     		node[sloped,inner sep=0cm,above,pos=\number\p*0.01,
								anchor=south east, minimum height=(10)*0.2cm,
								minimum width=(100-\p)*0.02cm+1cm] (N2 \p){}
								};
							\path  (2,0)--node[below=.5em] {}(0,0);
							\draw[wilsonline,Maroon] (1,2) -- (1.5,3);
							\draw[wilsonline,Maroon] (1.5,3)--(4.5,3);
							\draw[wilsonline,Maroon] (4.5,3)--(3,0);
							\draw[wilsonline,Maroon] (3,0)--(2,0);
			
							\node (E) at (1.5,3) {${\color{Maroon}\bullet}$};
							\node (F) at (4.5,3) {${\color{Maroon}\bullet}$};
							\node (H) at (3,0) {${\color{Maroon}\bullet}$};
								
							\node[right=.1cm] at (.5,4.5) {${\color{blue} V^\m(t)=(l_1^{+},l_2^{-},0^\perp)}$};	
							\foreach \p in {10,20,...,100} {
						      		\draw[-latex,blue,thick] (N \p.south west) -- (N \p.north east);
						    		}
							\foreach \p in {10,20,...,80} {
						      		\draw[-latex,blue,thick] (N2 \p.south east) -- (N2 \p.north west);
						    		}	
							\Looppoints					
							\end{scope}

						\end{tikzpicture}
			\caption{Generating variational vector field.}
			\label{fig:diffeofield}
		\end{figure}

			Taking the trace over the color matrices then adds the color factor $C_N$ and using
			the linearity of the wedge product in the vector field $V^\m$ we have the final result:
				\be
					\left(\ell_1\frac{ \delta}{\delta \ell_1} + \ell_2\frac{ \delta}{\delta \ell_2}\right) \  {\cal W}_\g  = D_V \ {\cal W}_\g  \ ,		
 		\ee
				with $V^\m = V_1^\m+V_2^\m = (\ell_1^{+} \ , \ \ell_2^{-} \ , \ \vecc 0_\perp)$ (see also figure \ref{fig:diffeofield}).
				Taking into account the renormalization properties of the light-like Wilson quadrilateral loop \cite{WL_LC_rect_1,WL_LC_rect_2,WL_LC_rect_3,Cherednikov:2012yd}, we come to our final result:
\begin{equation}
 \mu \frac{d}{d\mu} \ \left[ D_V \  {\cal W}_\g \right]
 =
- \sum \Gamma_{\rm cusp} \ ,
\label{eq:final}
\end{equation}
where $\Gamma_{\rm cusp}$ is the light-cone cusp anomalous dimension \cite{KR87,WL_LC_rect_1,WL_LC_rect_2,WL_LC_rect_3} and the summation runs over the number of cusps.

				
\section{Discussion and Outline}

After introducing classically the logarithmic Fr\'echet derivative as a diffeomorphism induced derivative with associated variational vector field $V^\m$ we have shown that its lowest order quantum field-theoretic contribution is equivalent to the derivative $\left(\ell_1\frac{ \delta}{\delta \ell_1} + \ell_2\frac{ \delta}{\delta \ell_2}\right) $ we introduced in Ref. \cite{Cherednikov:2012yd}.
Therefore, we demonstrated explicitly that an important class of ``motions'' (which apparently is not taken into account straightforwardly within the Makeenko-Migdal approach) in the generalized loop space can be described by using the mathematically consistently defined Fr\'echet derivative. Since diffeomorphisms cannot bring about new cusps, the number of cusps is diffeomorphism-invariant. We would expect then that the light-like Wilson polygonal loops having different number of cusps relate to different physical objects.

On the other hand, diffeomorphism-invariant transformations of the light-like loops find straightforward applications in the analysis of UV and rapidity evolution of gauge-invariant correlation functions.		
In particular, a useful duality relation exists between this class of the paths transformations in the generalised loop space and rapidity evolution of certain matrix elements. Namely, rapidities attributed to the light-like vectors $\ell_{1,2}$ are formally infinite:
\begin{equation}
 y_{1,2}
 =
 \frac{1}{2} \ \ln \frac{\ell_{1,2}^+}{\ell_{1,2}^-}
 \sim  \pm \frac{1}{2}
\lim_{\eta^\pm \to 0}  \ \ln \frac{(\ell_1 \cdot \ell_2)}{\eta^\pm} \ ,
\label{eq:rapidity}
\end{equation}
where $\eta^\pm$ is a regulator and plus- and minus- components of a vector $a_\mu$ are given by the scalar products $a^\pm = (a \cdot n^\mp)$ with $n^\mp \sim \ell_{2,1}$.
Eq. (\ref{eq:rapidity}) demonstrates, clearly, that
\begin{equation}
\frac{d }{d \ln S_{ij}} \sim \pm \frac{d }{d y_{i,j}}  \ .
\end{equation}
Validity of Eq. (\ref{eq:final}) in the higher orders of the perturbative expansion has been established recently \cite{CM_2014}. 
Our results suggest, therefore, that the rapidity evolution of a given correlation function is dual to the area transformations of a properly defined class of elements of the generalized loop space.

In particular, it has been demonstrated in \cite{ICh_FBS_2014} that the following factorization for the transverse-distance dependent parton density ${\cal F} \left(x, {\bm b}_\perp\right)$ is valid in the large Bjorken-$x$ approximation:
\begin{equation}
{\cal F} \left(x, {\bm b}_\perp; P^+, n^-, \mu^2 \right)
\approx
{\cal H} (\mu^2, P^2) \cdot {\Phi} (x, {\bm b}_\perp; P^+, n^-, \mu^2 ) \ ,
\label{eq:LargeX_factor}
\end{equation}
where the contribution the $x$-independent jet function ${\cal H}$ describes the incoming-collinear partons and 
the soft function $\Phi$ can be defined as the Fourier transform of an element of the generalized loop space
\begin{equation}
{\Phi} (x, {\bm b}_\perp; P^+, n^-, \mu^2 )
= \int\!dz^- \ {\rm e}^{-i (1-x) P^+ z^-} \ {\cal W}_\Pi (z^-, {\bm b}_\perp; P^+, n^-, \mu^2 ) \ ,
\label{eq:soft_LargeX}
\end{equation}
where the so-called {\it double-$\Pi$ shape} Wilson loop reads  
\be
{\cal W}_\Pi (z^-, {\bm b}_\perp; P^+, n^-, \mu^2 ) =
 \langle 0 | \  {\cal U}_P^\dag [\infty; z] {\cal U}_{n^-}^\dag[z; \infty] {\cal U}_{n^-} [\infty ; 0] {\cal U}_P [0; \infty]
  \  | 0 \rangle
\ee
which contains the following Wilson lines: incoming-collinear (off-light-cone, $P^2 \neq 0$), ${\cal U}_P$, and outgoing-collinear (light-like, $(n^-)^2 = 0 $), ${\cal U}_{n^-}$.
Therefore, the pure quark effects (accumulated in ${\cal H}$) get separated out from the soft part $\Phi (x, {\bm b_\perp})$, which contains complete information about the three-dimensional structure of the nucleon in the large-$x_B$ regime accessible at the EIC and JLab. 

On the other hand, this is the soft part that determines the rapidity evolution of the whole function (\ref{eq:LargeX_factor}). Therefore, our result on the connection between diffeomorphism-invariant transformations in the loop space and classically conformal invariant shape variations imply that the calculation of the evolution kernels can be made simpler within this approach. Namely, the rapidity evolution of a certain Wilson loop can be re-written in terms of the appropriate Fr\'echet derivative, which allows one to derive the complete set of the rapidity-ultraviolet evolution equations. This result will be reported elsewhere.

\section{Acknowledgements}

We appreciate the numerous fruitful discussions in the course of this work with Frederik Van der Veken and Pieter Taels.

\end{document}